\documentclass[intlimits,twoside,a4paper]{article}

\usepackage{graphicx,psfrag}

\usepackage[T2A]{fontenc} %T2A
\usepackage[utf8]{inputenc} %cp1251
\usepackage[ukrainian,english]{babel}

\usepackage[eqsecnum]{cmpj3}

\RequirePackage{color}

\issue{2021}{24}{3}{33603}
\doinumber{10.5488/CMP.24.33603}

\title[DNA thermal denaturation by polymer field theory approach:  effects of the 
environment]%
{DNA thermal denaturation by polymer field theory approach:  effects of the 
environment}
\author[Yu. Holovatch, C. von Ferber, Yu. Honchar]{Yu. Holovatch\orcid{0000-0002-1125-2532} 
\refaddr{label1,label2,label3}, 
C. von Ferber\orcid{0000-0002-6240-9237} 
\refaddr{label2,label3},  Yu. Honchar\orcid{0000-0003-2660-4593} 
\refaddr{label1,label2,label3}\thanks{Corresponding author: \email{julkohon@icmp.lviv.ua}}
}
\addresses{
\addr{label1}Institute for Condensed Matter Physics of the National
Academy of Sciences of Ukraine,\\ 1 Svientsitskii St., 79011 Lviv,
Ukraine 
\addr{label2}$\mathbb{L}^4$ Collaboration \& Doctoral College for the Statistical Physics of Complex Systems, 
Leipzig-Lorraine-Lviv-Coventry, Europe 
\addr{label3}Centre for Fluid and Complex Systems, Coventry University, Coventry, 
CV1 5FB, United Kingdom}

\date{Received June 9, 2021, in final form July 19, 2021}
\begin{document}
\maketitle
\begin{abstract}
We analyse the effects of the environment (solvent quality, 
presence of extended structures --- crowded environment) that may have impact on the order of 
the transition between denaturated and bounded DNA states and lead to
changes in the scaling laws that govern conformational properties of
DNA strands.
%We show that different environments may shift the transition towards or
%away from the first order regime. 
We find that the effects studied 
significantly influence the strength of the first order transition.
To this end, we re-consider the Poland-Scheraga model and
apply a polymer field theory to calculate entropic exponents associated
with the denaturated loop distribution. For the $d = 3$
case, the corresponding diverging $\varepsilon=4-d$ expansions are
evaluated by restoring their convergence via the resummation technique.
 For the space dimension $d = 2$,
the exponents are deduced from mapping the polymer model onto a two-dimensional
random lattice, i.e., in the presence of quantum gravity.
We also show that the first order transition is further strengthened by the
presence of extended impenetrable regions in a solvent that restrict the number of 
the macromolecule configurations.
\printkeywords
%
%\keywords  DNA denaturation, Poland-Scheraga model, polymer networks, crowded environment.  
%
%\pacs 64.60.aq, 64.60.fd, 64.70.qd, 64.60.De
\end{abstract}

Nucleic acids together with proteins and carbohydrates belong to
macromolecules essential to all known forms of life. 
Enormous experimental, theoretical, and simulational efforts
have been involved to understand and qualitatively describe their 
physical, chemical and biological properties. In this paper
we show how an insight from polymer field theory helps
to shed light on properties of a DNA helix-to-coil (also called 
denaturation,  unwinding or unzipping) transition: a
phenomenon, that lies at the origin of biological processes involving
DNA, as duplication or transcription. The latter phenomena occur
in a cell and are complex biological protein-mediated processes. 
An analogue of DNA unwinding is also observed {\em in vitro}:
when purified DNA solution is heated above the room temperature, 
the  cooperative transition from  the hydrogen bound double-stranded 
helix structure to a single stranded one occurs, see~\cite{Wartell85} and references
therein for review. This phenomenon is known as DNA thermal denaturation and
is the subject of our study. 

In statistical physics, the DNA thermal denaturation is described in terms of 
the Poland-Scheraga model \cite{Poland66,Poland66a,Poland70} that allows its treatment 
in terms of phase transition theory.
In a recent paper \cite{Honchar21a} we have shown that changes in the solvent
quality may cause an essential impact on the order of the phase transition between
denaturated and bounded DNA states. To quantify this impact, we have calculated
 $\varepsilon=4-d$ expansions for the 
entropic exponents that govern the denaturated loop distribution in a good solvent
and in the $\theta$-solvent regimes and evaluated these (divergent) expansions in $d=3$.
In this paper, we complement such analysis by offering exact results for the exponents
at $d=2$. Moreover, we further analyse possible reasons that may have impact on the order
of the transition. In particular, we are interested in the effects caused by the presence of 
extended structures that restrict the swelling of polymer chains. By such analysis
we make an attempt to consider the situation in a more realistic condition of
macromolecules in a crowded environment of a cell \cite{Reiter15}. 

\begin{figure}[!t]
\centering\includegraphics[]{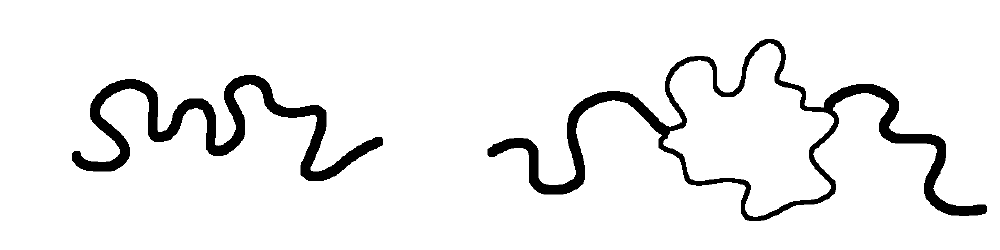}\\
\centering{{\bf a.} \hspace{15em}  {\bf b.}}
\caption{Model of DNA thermal denaturation (unzipping) transition discussed in this
paper. At low temperatures $T$, two DNA strands remain connected by hydrogen bonds and form a 
single long flexible polymer chain (figure~{\bf a}). With an increase of $T$, the chain
unzips and a loop emerges: now the whole structure consists of two different `species':
double stranded side chains, shown by solid lines, and a single stranded loop shown by a
thin line in the middle (figure~{\bf b}).
\label{fig1}}
\end{figure}

The rest of the paper is
organized as follows. In the next section we repeat some of our results for the
scaling relations and entropic scaling exponents \cite{Honchar21a} and give
their numerical estimates at $d=3$. 
Based on the exact conformal scaling dimensions for two-dimensional copolymers 
derived from an algebraic structure existing on a random lattice (quantum gravity)
\cite{Duplantier99,Duplantier99a,vonFerber04} we derive
exact values of the exponents at $d=2$  and discuss the whole sector 
$2\leqslant d \leqslant 4$ in section \ref{III}. Crowded environment effects are analysed in
section \ref{IV}, conclusions are summarized in section \ref{V}.
It is our pleasure and a big honour to contribute by this paper to the Festschrift
devoted to Prof. Yu. Kalyuzhnyi on the occasion of his 70th birthday. Doing so we deeply
acknowledge his seminal contributions to the soft matter physics in general and to the
subject discussed in this paper in particular, see e. g., \cite{Kalyuzhnyi,Kalyuzhnyi1,Kalyuzhnyi2,Kalyuzhnyi3,Kalyuzhnyi4,Kalyuzhnyi5}. CvF and YuH
also are indebted to the jubilee for a long-lasting friendship, numerous discussions
about physics and not only.

\section{Poland-Scheraga model: scaling relations and $\varepsilon$-expansion}\label{I}
The model suggested by Poland and Scheraga in middle-sixties \cite{Poland66, Poland66a}
describes the DNA thermal denaturation by a proper account of energy-entropy interplay: at low temperatures
$T$, the bound state, figure~\ref{fig1} {\bf a}, is favoured by energy whereas
at high $T$  the unbound state, figure~\ref{fig1} {\bf b}, is favoured by entropy as
the one having more configurations. 
Poland and Scheraga's theoretical works lead to a whole family of DNA denaturation 
models \cite{Poland70,Reiter15,Fisher66,Causo00,PSnetwork,PSnetwork1,PSnetwork2,PSnetwork3,PSmath,PSmath1,PSmath2}.
It was shown that the unzipping transition mechanism is governed by the universal loop exponent $c$ which describes
scaling of the partition function of a single-stranded DNA loop in  double stranded side chains, see figure~\ref{fig1} {\bf b}:  
\begin{equation} \label{I_1}
	{\cal Z}_{\rm loop} \sim \mu^\ell \ell^{-c},
\end{equation}
here, $\ell$ is loop length (number of unbound segments) and $\mu$ is non-universal fugacity. For $c>1$, the model
predicts the denaturation transition whereas for  $0 \leqslant c \leqslant 1$ the order parameter (average
number of ordered bound pairs in a chain) is a continuous function of $T$ smoothly changing between 1 and 0 when
$T$ increases from 0 to $\infty$. In turn, for larger values of $c$, the order parameter either continuously
vanishes at $T=T_c$ for $1<c\leqslant 2$ or disappears abruptly at $T=T_c$ for $c > 2$.
The last two types of behaviour correspond to the second and first order phase transitions, respectively.
However, the value of $c$ is not obvious. First papers on the model suggested $c=d/2$, which lead to the second order transition 
and $d=3$~\cite{Poland66, Poland66a}. Later Fisher has considered taking into account the self-avoiding nature of chains that
lead to $c = d\nu$~\cite{Fisher66}, where $\nu$ is polymer mean square end-to-end distance scaling exponent. Still, with $\nu(d=3)\simeq 0.588$ ~\cite{desCloizeaux91}, the phase 
transition remains the second order. This result contradicts experimental observations of the
first order nature of the transition  \cite{Wartell85}. A more general approach to analyze scaling properties of the 
macromolecule configurations shown in 
figure~\ref{fig1} was based on polymer network theory, as interaction between the loop and the chain was taken into account 
\cite{Causo00,PSnetwork,PSnetwork1,PSnetwork2,PSnetwork3,Baiesi02,Blossey03}. Considering both the side chains and the loop as self-avoiding walks (SAWs), 
it was shown that the phase transition 
is of the first order for $d=2$ and above. This result was further supported by numerical simulations \cite{Baiesi02} and it was also suggested
that possible heterogeneity in chain structure may strengthen the transition.

\begin{figure}[!t]
\centering\includegraphics[]{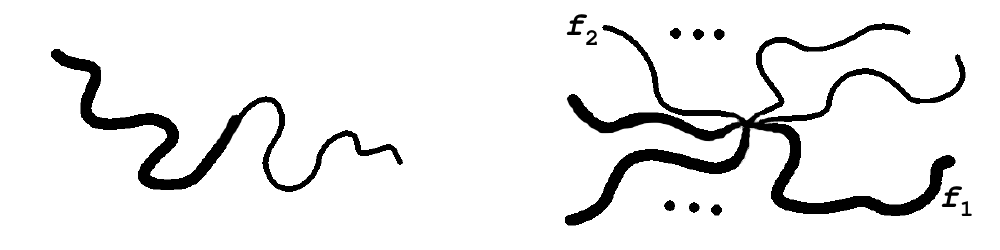}\\
\centering{{\bf a.} \hspace{15em}  {\bf b.}}
\caption{{\bf a:} block copolymer of two polymer chains
of different species, shown by solid and thin lines, linked together.
{\bf b:} copolymer star consisting of $f_1$ chains species 1
and $f_2$ chains of species 2 tied together at their end points. Its scaling properties
are governed by universal copolymer star exponents $\eta_{f_1f_2}$. Note that the block copolymer
gives a trivial example of a two-arm copolymer star with an exponent $\eta_{11}$.
\label{fig2}}
\end{figure}

Depending on temperature, the asymptotic scaling behaviour of a long flexible polymer macromolecule 
in a good solvent belongs either to random walk (RW), $T=T_\theta$, or to self-avoiding walk (SAW), $T>T_\theta$ universality 
classes ($T_\theta$ denoting the $\theta$-point) \cite{ternary,ternary1}. Therefore, the only difference that may be observed 
in asymptotic scaling of chains of different species (in our case these are the double- and single-stranded chains) is
due to the difference in asymptotic scaling properties of mutually
interacting SAWs and RWs. Based on this fact, recently \cite{Honchar21a} we have applied 
polymer field theory \cite{vonFerber04,desCloizeaux91, vonFerber97,vonFerber97a,vonFerber97b}
to derive scaling relations that express
the loop exponent $c$ (\ref{I_1}) in terms of the familiar copolymer star exponents~$\eta_{f_1f_2}$.  
The latter govern the scaling of star-like polymer structures
that are created by linking together the
end points of polymer chains of two different species at a common core, as shown in figure~\ref{fig2}. 
When such a copolymer star is immersed in a good solvent,
its asymptotic properties are universal in the limit of long
chains. In particular, the partition function (the number of configurations) 
of a copolymer star made of two sets of $f_1$ and $f_2$ mutually avoiding RWs
scales with its size $R$ as  \cite{vonFerber04,vonFerber97,vonFerber97a,vonFerber97b}:
\begin{equation}\label{I_1a}
 Z^G_{f_1f_2} \sim R^{\eta^G_{f_1f_2}}\, .
\end{equation}
In turn, the partition function 
of a copolymer star made of mutually avoiding sets of $f_1$ SAWs and $f_2$ RWs
scales as:
\begin{equation}\label{I_1b}
 Z^U_{f_1f_2} \sim R^{\eta^U_{f_1f_2} - f_1\eta^U_{20}}\, .
\end{equation}
The third case which is of interest here is the star of two sets of $f_1$ and $f_2$ SAWs.
For its partition function, one gets:
\begin{equation}\label{I_1c}
 Z^S_{f_1f_2} \sim R^{\eta^S_{f_1,f_2} - (f_1+f_2)\eta^S_{20}}\, .
\end{equation}
Indices $G,U,S$ in the above formulae refer to the fixed points (FPs) of the renormalization
group transformation that govern the scaling of corresponding mutually avoiding structures: Gaussian FP for
RWs, unsymmetric FP for RW and SAW, and symmetric FP for SAWs, see \cite{vonFerber97,vonFerber97a,vonFerber97b} for more details.
Exponents $\eta^S_{f_1,f_2}$ are related to $\eta^U_{f_1,f_2}$ and to the homogeneous star exponents
$\eta_f$ \cite{stars,stars1} via: $\eta^S_{f_1,f_2}=\eta^U_{f_1+f_2,0}=\eta_{f_1+f_2}$.

With the above considerations in mind, one is led to four different cases that account 
for possible inhomogeneities and, therefore, for different scaling exponents 
of the DNA denaturation model shown in figure~\ref{fig1} {\bf b}: 
\begin{enumerate}
	\item both bound chains and the unbound loop are SAWs (SAW-SAW-SAW);
	\item bound chains are SAWs, the loop is RW (SAW-RW-SAW);
	\item the chains are RW-like, while the loop is SAW (RW-SAW-RW);
	\item both the chains and the loop are RW-like, though they do not
	intersect each other (RW-RW-RW).
\end{enumerate}
The scaling relations that express the loop exponent  $c$ (\ref{I_1}) in terms of copolymer
star exponents $\eta_{f_1f_2}$ for these four cases read \cite{Honchar21a}:
\begin{align} \label{I_2}
&\text{1.  SAW-SAW-SAW:} &c_1 = \nu_{\rm SAW} (3\eta^S_{20} + d - 2\eta^S_{12})\, , \\ \nonumber
&\text{2. SAW-RW-SAW:} &c_2 = \nu_{\rm RW} (\eta^S_{20} + d - 2\eta^U_{12})\, , \\ \nonumber
&\text{3.  RW-SAW-RW:} &c_3 =  \nu_{\rm SAW} (2\eta^S_{20} + d - 2\eta^U_{21})\, , \\ \nonumber
&\text{4.  RW-RW-RW:} &c_4 = \nu_{\rm RW} (d - 2\eta^G_{21})\, .
 \end{align}
Here, $\nu_{\rm RW} = 1/2$ and $\nu_{\rm SAW}$ are the mean square end-to-end distance exponents for the random and self-avoiding walks,
correspondingly, and $d$ is space dimension. The exponents $\eta_{f_1f_2}$ have been calculated within field-theoretical
renormalization group approach \cite{vonFerber04,vonFerber97,vonFerber97a,vonFerber97b} and are currently know in the fourth order 
of the $\varepsilon=4-d$ expansion \cite{Schulte-Frohlinde04}.
Below, we list them together with the  $\varepsilon$-expansion for the exponent $\nu_{\rm SAW}$ \cite{Kleinert}
in the corresponding order:
\begin{align}\label{eta1}
	\eta^S_{20} (\varepsilon) =&- \varepsilon/4 - 9\varepsilon^2/128+ \varepsilon^3 [264\zeta(3) - 
	49]/2048\\ &+\varepsilon^4[704\piup^4 - 297600\zeta(5) + 38160\zeta(3) + 235]/655360 \, ,\nonumber
\end{align}
\begin{align}\label{eta2}
	\eta^S_{12} (\varepsilon) =&  - 3\varepsilon/4- 3\varepsilon^2/128+ 3\varepsilon^3[40\zeta(3) + 23] /2048  \\ &+\varepsilon^4[64\piup^4 - 32640\zeta(5) - 6480\zeta(3) + 3333]/131072\,,\nonumber
\end{align}
\begin{align}\label{eta3}
\eta^U_{12} (\varepsilon) =& - 3\varepsilon/4 + \varepsilon^2[42\zeta(3) - 13]/128 + \varepsilon^3[384\zeta(3) - 5]/2048  \\&+\varepsilon^4[1024\piup^4 - 528000\zeta(5) + 14880\zeta(3) + 7655]/655360\,,\nonumber
\end{align}
\begin{align}\label{eta4}
	 \eta^U_{21} (\varepsilon) = &- \varepsilon + \varepsilon^2[42\zeta(3) + 1]/64+ 17\varepsilon^3/1024   \\ &-\varepsilon^4[1056\zeta(3) - 721]/65536\,, \nonumber
\end{align}
\begin{equation}\label{eta5}
	\eta^G_{21} (\varepsilon) = - \varepsilon ,
\end{equation}
\begin{align}\label{nu}
	\nu_{\rm SAW} (\varepsilon) =&1/2 +\varepsilon/16+15\varepsilon^2/512+\varepsilon^3[135/8192-(33/1024)\zeta(3)]  +\varepsilon^4[3799/524288\\
	&-(873/32768)\zeta(3)-(11/40960)\piup^4+(465/4096)\zeta(5)] \nonumber  \, ,
	\end{align}
where $\zeta(x)$ is Riemann zeta-function. Note that the formula for the exponent $\eta^G_{21}$ contains 
only linear in $\varepsilon$ term and is exact.

Substituting expressions (\ref{eta1})-(\ref{nu}) into the scaling relations (\ref{I_2}) one can evaluate loop
exponents $c_i$ at any value of $d$. It is well known, however, that $\varepsilon$-expansions
of the field theory are asymptotic at best and proper resummation technique is required to get a reliable
numerical information on their basis \cite{Hardy48,Zinn-Justin89}. Applying resummation technique based on the Borel-Leroy
transformation enhanced by conformal mapping of a cut-plane on a disc \cite{LeGuillou80,Delamotte,Delamotte1}, we arrived
at the following values of the loop exponents $c_i$ for $d=3$ \cite{Honchar21a}:
\begin{eqnarray} \label{II_c}
	c_1&=&2.147 \pm 0.009, \hspace{1em}
	c_2=2.169 \pm 0.004, \\ \nonumber
	c_3&=& 2.76 \pm 0.03, \hspace{2em} 
	c_4= 2.5  .
\end{eqnarray}
Clearly, $c>2$ in all configurations, which confirms the first order transition.
\begin{figure}[!t]
	\centering\includegraphics[height=6.5cm]{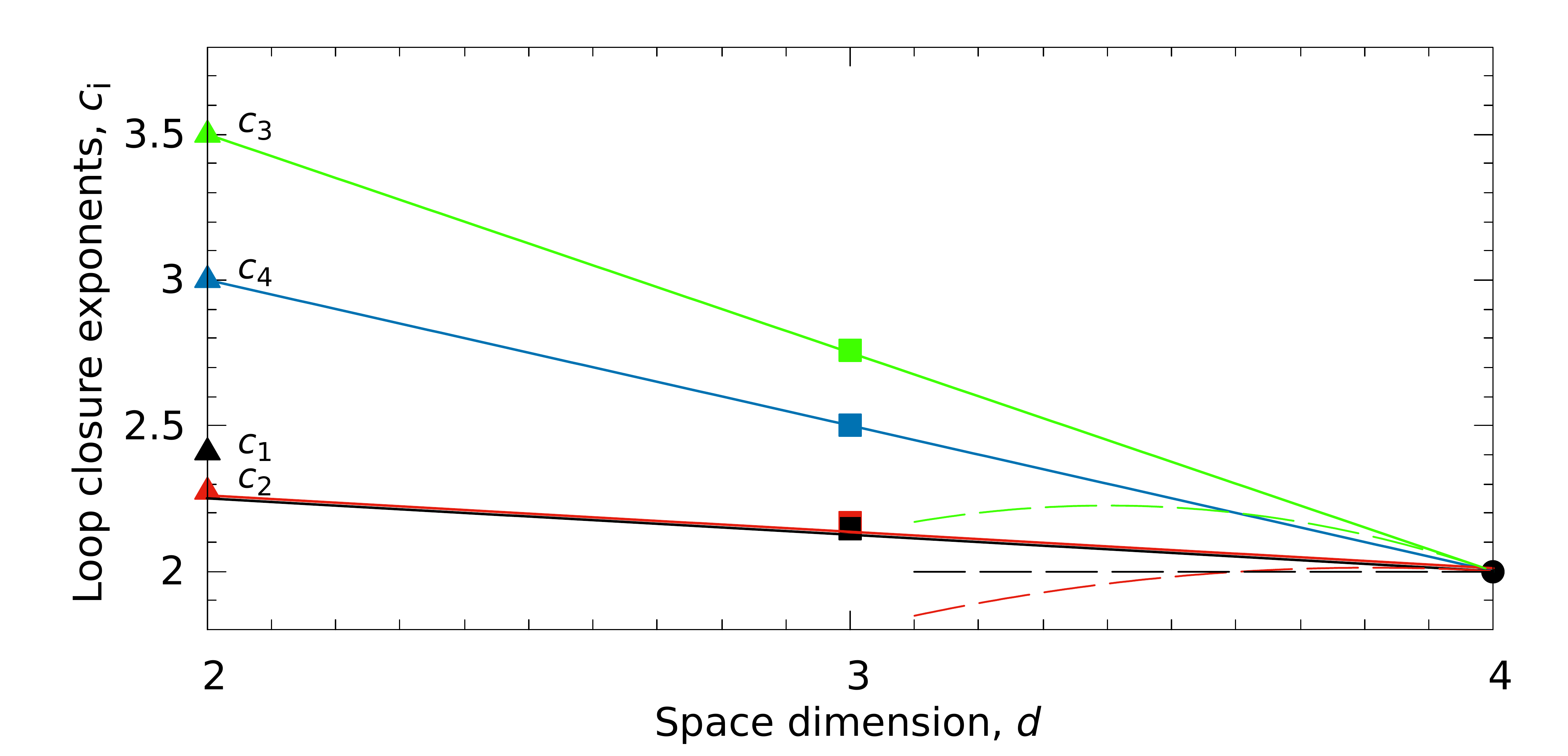}
	\caption{(Colour online) Loop closure exponents $c_i$ at different space dimension $d$.
		Triangles show exact results at $d=2$ (\ref{III_3}) and squares show most accurate results
		obtained by resummation at $d=3$ \cite{Honchar21a}. The lines show analytic continuation
		to non-integer $d$ via $\varepsilon$-expansion in the first and second orders,
		thick solid and thin dashed lines, correspondingly.
		Note that in the case of mutually avoiding RWs (blue line), loop closure exponents $c_4(\varepsilon)$ are exact
		and linear in $\varepsilon$.
		\label{fig3}}
\end{figure}

\section{Exact results at $d=2$ and quantum gravity}\label{III}
As has been discussed above, the $\varepsilon$-expansions for the
$c_i$ may serve as a basis for reliable numerical estimates at $d=3$
provided appropriate resummation technique is applied. With the
perturbative expansions  and their numerical estimates at hand,
it is instructive to corroborate the results by comparing them
with the data for other space dimensions, if available. One obvious
result is obtained for $d=4$: there, as it is easy to check via equations (\ref{I_2}),
all exponents are equal: $c_i(d=4)=2$.
Besides, there is a tempting opportunity to get exact values
for the exponents at $d=2$. Indeed, to this end one can make
use of the exact results for the scaling exponents of $d=2$ copolymer 
stars of mutually avoiding bunches of SAW and RW \cite{Duplantier99,Duplantier99a}.
There, the relations between exponents in fluctuating geometry 
(quantum gravity) and flat $d=2$ geometry have been used to extract
the exact values of the exponents. In notations of the previous section,
the exponents read:\footnote{Cf. equations (100), (101) of \cite{vonFerber04}.}
\begin{eqnarray} \label{III_1}
\eta^G_{f_1f_2}&=&\frac{1}{48}\Big \{4 - \Big [\sqrt{24f_1+1} + \sqrt{24f_2+1}-2 \Big ]^2 \Big \} \, ,
\\ \label{III_2}
\eta^U_{f_1f_2}&=&\frac{1}{48}\Big \{ 4 +5f_1 - \Big [3f_1 + \sqrt{24f_2+1} -1 \Big ]^2 \Big \} \, .
\end{eqnarray}
Substituting these formulae into equations (\ref{I_2}) and taking into account that 
$\nu_{\rm SAW}(d=2)=3/4$ \cite{desCloizeaux91}, one gets the following exact values 
of the exponents $c_i$ at $d=2$:
\begin{eqnarray} \label{III_3}
c_1&=&\frac{77}{32}\simeq 2.406, \hspace{1em}
c_2=\frac{109}{48}\simeq 2.271, \\ \nonumber
 c_3&=& \frac{7}{2}, \hspace{1em}
 c_4= 3 \, .
\end{eqnarray}
These values are shown by triangles in figure~\ref{fig3}. The obtained result for the exponent $c_3$
recovers the value predicted at $d=2$ by the exact formula that follows from equation (\ref{I_2}) 
and is also valid  for other values of $d$:
$c_3=2+ \varepsilon/2$. 

Comparing the values of the loop closure exponents $c_i$ at $d=2$ and at $d=3$ one can arrive at 
certain conclusions about an impact of chain heterogeneity on the strength of the DNA thermal
denaturation transition. The first observation is that passing from the homogeneous SAW composition
within the Poland-Scheraga model (as described by the exponent $c_1$) usually leads to 
strengthening of the first order transition. When the SAW side chains
are substituted by RWs, the strength of the transition increases: $c_3>c_1$, $c_4>c_2$. In turn,
when the side chains remain unchanged, the change of the SAW loop to the RW loop decreases the strength of
the first order transition: $c_3>c_4$ and $c_1>c_2$. The last effect is more pronounced for the RW side chains
and at space dimension $d=2$. In general, the following rule holds: $c_2<c_1<c_4<c_3$ (with $c_2\sim c_1$ at $d=3$).

Another striking feature that follows from the comparison of the exact and perturbative results shown in figure
\ref{fig3} is a rather unusual behaviour of the $\varepsilon$-expansion curves. Indeed, the first order $\varepsilon$-expansion
for the exponents $c_i$ (solid lines in the plot) nicely corresponds to the resummed $\varepsilon^4$-data
at $d=3$ and to the exact values at $d=2$. Such a behaviour is obvious for the exponent $c_3$, where the first order
$\varepsilon$-expansion provides an exact number. However, for the other exponents, an account of the higher orders of
the perturbative expansion needs careful application of the resummation technique. Being evaluated na\"ively by simple
adding higher order contribution, the $\varepsilon$-expansion holds only very close to the upper critical
dimension $d=4$, as shown in the figure by the thin dashed lines for the case of $\varepsilon^2$-data. Therefore, the first
order $\varepsilon$-expansion provides the so-called optimal truncation \cite{Hardy48} for the $c_i(\varepsilon)$ series.

\section{Crowded environment}\label{IV}

In two former sections, \ref{I} and \ref{III}, we discussed an impact of the solvent quantity 
on the order of the DNA thermal denaturation transition. Another factor
that may modify the scaling exponents of long flexible polymer macromolecules  is the 
presence of impurities --- impenetrable regions in a solvent that restrict the number of polymer
configurations, see e.g., \cite{Chakrabarti05} and references therein. Statistics of 
polymers in disordered medium is of interest for a number of reasons. In the context
of our study it is important to mention its relevance for treating macromolecules
in a cell, composed of many different kinds of biochemical species \cite{cell,cell1,cell2}.

There exist different analytic frameworks to model an impact of disordered medium on the
scaling properties of (interacting) SAWs and RWs. To give a few examples, the latter
are studied on a percolation cluster \cite{percolation,percolation1} or at presence of quenched
defects \cite{short-range,short-range1,Weinrib83,Blavatska13,Blavatska01,Blavatska01a,Blavatska01b,Blavatska11}. Taking into consideration that 
the uncorrelated defects do not influence polymer
scaling \cite{short-range,short-range1}, the so-called `extended' or long-range correlated 
structural disorder has been shown to be relevant. A model of long-range
correlated disorder has been suggested in  \cite{Weinrib83} and further
exploited in studies of polymers \cite{Blavatska01,Blavatska01a,Blavatska01b,Blavatska11}. Within this model, one considers the 
defects, characterized by the density-density pair correlation function $g(r)$ decaying
at a large distance $r$ according to the power law
\begin{equation}\label{IV_1}
g(r) \sim r^{-a}\, .
\end{equation}
For integer values of $a$, such defects have a direct interpretation:
the case $a=d$ corresponds to point-like defects, while $a=d-1$ ($a=d-2$) correspond to straight 
lines (planes) of defects of random orientation. Sometimes non-integer values of $a$ are interpreted 
in terms of fractal structures.\footnote{See also  \cite{fractals,fractals1,fractals2}, where
the relation of fractal dimension to the analytically continued non-integer dimension
is discussed in more details.} Detailed analysis of an impact of the long-range correlated
disorder (\ref{IV_1}) on possible changes in the exponents (\ref{I_2}) and hence on the DNA thermal
denaturation is beyond the scope of this study. However, we will use some of the
previously obtained results in order to understand and qualitatively describe this
possible impact. 

It is easy to see that the presence of long-range correlated impurities may or may not be relevant 
and change the polymer scaling exponents depending on the value of $a$. Indeed, large-distance 
asymptotics of the pair correlation function (\ref{IV_1}) corresponds to the power-law behaviour 
of its Fourier-image  at small wave vector $k$ in the form $k^{a-d}$. Therefore, by simple 
power counting, one arrives at the conclusion that such a term becomes relevant at small $k$ 
for $a<d$. Applying field-theoretic renormalization group technique, the corresponding polymer
model has been analysed and the scaling exponents were calculated in the two-loop approximation 
at fixed $d=3$ and different values of the correlation parameter $a$ as well as in a
one-loop order by the double expansion in $\varepsilon=4-d$ and $\delta=4-a$
\cite{Blavatska01,Blavatska01a,Blavatska01b}. The derivation given below is based on these double
$\varepsilon,\delta$ expansions. In particular, it has been shown that for certain
region of parameters $\varepsilon/2 < \delta < \varepsilon$, the scaling properties of a 
single flexible polymer chain in porous environment with a long-range correlated
structure are governed by a new, `long-range'
fixed point $L$. The mean square end-to-end 
distance  exponent $\nu_{\rm SAW}$ in the first order of
$\varepsilon,\delta$ expansion reads \cite{Blavatska01}:
\begin{equation} \label{IV_2}
\nu^L_{\rm SAW} = 1/2 + \delta/8 + \dots \, .
\end{equation}
In turn, the $\eta_{f_1f_2}$ exponents  for co-polymer stars  in 
porous environment with long-range correlated structure are
given by:\footnote{Cf. equation (39)
from  \cite{Blavatska11}.}
\begin{eqnarray} \label{IV_3}
\eta^{S_L}_{f_1f_2} &=& \frac{-(f_1+f_2)(f_1+f_2-1)\delta}{4} \, ,\\
\label{IV_4}
\eta^{U_L}_{f_1f_2} &=& \frac{-f_1(f_1+3f_2-1)\delta}{4} \, ,\\
\label{IV_5}
\eta^{G_L}_{f_1f_2} &=& -f_1f_2\delta \, .
\end{eqnarray}
In equations (\ref{IV_3})--(\ref{IV_5}), the first exponent $\eta^{S_L}_{f_1f_2}$ corresponds to the star of
$f_1+f_2$ SAWs,
the second exponent $\eta^{U_L}_{f_1f_2}$ describes the star of mutually avoiding sets of $f_1$ SAWs and $f_2$ RWs,
and the third exponent $\eta^{G_L}_{f_1f_2}$ describes the star of two mutually avoiding sets of $f_1$ and $f_2$ RWs. 

Two cautions are at place here. First, the  `long-range'
fixed point $S_L$ is accessible in the region where the above
mentioned power counting shows that the disorder is irrelevant.
Second, the fixed points $U_L$ and $G_L$ can be reached only
for specific initial conditions. Similar situation is also encountered
 when the $\varepsilon,\delta$ expansion is applied
to study models of $m$-vector magnets with long-range correlated 
quenched disorder \cite{Weinrib83}. However, an account of
higher order contributions restores the physical region of stability
of the `long-range' fixed point confirming a qualitatively correct
result of the first-order analysis, see e.g., \cite{Holovatch02} and
references therein. Therefore, with an aim of getting a qualitative
description of an impact of extended long-range correlated
impurities on the DNA thermal denaturation transition, we proceed
with formulae (\ref{IV_2})--(\ref{IV_5}) substituting them
into the scaling relations (\ref{I_2}) and arrive at the following 
first-order values for the $c_i$ exponents:
\begin{eqnarray} \label{IV_6}
c^L_1&=&c^L_2=2-\varepsilon/2+5\delta/4 \, , \\ \label{IV_7}
c^L_3&=&c^L_4=2-\varepsilon/2+2\delta  .
\end{eqnarray}
As it follows from equation (\ref{IV_1}), the smaller is $a$, the stronger are the correlations in
porous structure that restricts the volume available for the macromolecule.
Indeed,  the density-density correlation function $g(r)$ decays
slower with a decrease of $a$, attaining the fat-tail features. The positive
sign at the linear in $\delta$ terms in equations (\ref{IV_6}), (\ref{IV_7}) brings about an 
increase in the exponents $c_i$ with an increase of $\delta=4-a$. This allows to conclude,
that an increase in 
density correlations of the porous structure leads to strengthening of the DNA 
thermal denaturation transition. Moreover, comparing equations (\ref{IV_6}) and (\ref{IV_7}), 
one concludes that $c^L_3, c^4_4 > c^L_1, c^L_2$, similar to what was observed for the 
DNA denaturation in a pure solvent without porous medium. The difference between the exponents
increases with an increase of $\delta$: $c^L_{3,4} -c^L_{1,2}=3\delta/4$. Of course,
with all cautions mentioned above, these results should be considered as  qualitative
predictions, rather than a quantitative description of DNA denaturation in a crowded
environment. The above obtained  relations
$c^L_1=c^L_2$ and $c^L_3=c^L_4$ may be (and perhaps indeed are)
violated in the second order of the perturbation theory.
However, it is worth mentioning that the scaling arguments supported by the 
renormalization group calculations predict the effect of strengthening the order of the
denaturation transition when it occurs in presence of extended structures that
restrict the swelling of the polymer coil.

\section{Conclusions}\label{V}

The value of the loop closure exponent $c$ (\ref{I_1}) discriminates between different ways the
thermal denaturation of the DNA occurs: for $c>2$, the denaturated loop emerges abruptly, in the
first order phase transition manner, for $1<c<2$, the transition is continuous, and
for $c<1$, no transition happens. Numerous attempts of theoretical description and
numerical simulation of this phenomenon finally led to the coherent picture, observed
also in the {\em in vitro} experiments and simulations \cite{Wartell85,Blake99,Causo00,Blossey03}:  
the transition is of the
first order and $c>2$. Besides, the factors that may have an impact on the strength, and, eventually,
even on the order, of this transition are discussed in the literature \cite{Reiter15}. In a recent 
paper~\cite{Honchar21a}, we have derived scaling relations that express the loop
closure exponent $c$ of the Poland-Scheraga model in terms of the copolymer star exponents
$\eta_{f_1f_2}$ \cite{vonFerber97,vonFerber97a,vonFerber97b}. This enabled us to analyse an impact of inhomogeneities in DNA
chain composition and solvent quality on the order of the transition. As it has been shown
in~\cite{Honchar21a}  and as it is briefly discussed in the above section~\ref{I}, consideration
of the macromolecule as sets of mutually avoiding SAWs and RWs (see figure~\ref{fig1}) leads to
an increase in value of $c$ and $d=3$ and, hence, strengthens the first order transition. 
In the present paper, we support this observation providing exact results at $d=2$. Moreover, we
show that the effect of strengthening is further enhanced by the so-called crowded environment
with the long-range correlated inhomogeneities.

\section{Acknowledgement} 

We acknowledge useful discussions with Maxym Dudka, Ralph Kenna, Mariana Krasnytska, and Dmytro Shapoval. 
This work was supported in part by the National Academy of Sciences of Ukraine, project KPKBK 6541230.

%\label{last@page}
%\end{document}
\pagebreak
\ukrainianpart

\title{Термічна денатурація ДНК в підході теорії поля для полімерів: вплив середовища}
\author{Ю. Головач\refaddr{label1,label2,label3}, К. фон Фербер\refaddr{label2,label3}, Ю. Гончар\refaddr{label1,label2,label3}}
\addresses{
	\addr{label1} Інститут фізики конденсованих систем Національної академії наук України, \\вул. Свєнціцького, 1, 79011 Львів, Україна
	\addr{label2} Співпраця $\mathbb{L}^4$ і Коледж докторантів `Статистична фізика складних систем', Ляйпціґ-Лотарингія-Львів-Ковентрі, Європа
	\addr{label3} Центр плинних і складних систем, Університет Ковентрі, Ковентрі, CV1 5FB, Великобританія
}
%
%% якщо автор є один або автори є з однієї установи:
%
%  \author{1й Автор, 2й Автор, \ldots}
%  \address{Інститут\ldots}
%
%%

\makeukrtitle

\begin{abstract}
	Ми розглянули вплив середовища (якість розчинника, присутність витягнутих структур (перешкод) --- «зайняте» середовище), який може змінити рід переходу між денатурованим та зв'язаним станами ДНК і привести до змін законів скейлінґу для конформаційних 
	властивостей ланцюжків ДНК. 
%	Ми показали, що різні середовища можуть зсувати цей перехід в напрямку поведінки переходу першого роду, 
%	або проти нього. 
	Показано, що досліджені ефекти значним чином впливають на інтенсивність переходу першого роду. 
	З цією метою, ми розглянули модель Поланда-Шераги і застосували підхід теорії поля для полімерів, щоби обчислити ентропійні 
	показники, пов'язані з розподілом денатурованих петель на ланцюгу. Для випадку $d = 3$ 
	проаналізовано відповідні розбіжні $\varepsilon = 4 - d$ розклади, оцінюючи їх за допомогою відновлення збіжності 
	методами пересумовування степеневих рядів.  Для вимірності $d = 2$ їх обчислено завдяки проектуванню 
	полімерної моделі на двовимірну випадкову ґратку, тобто розглянуто систему за присутності квантової ґравітації. 
	Ми також показуємо, що інтенсивність переходу першого роду посилюється за наявності у розчиннику протяжних непроникних областей, що 
	обмежують кількість конфігурацій макромолекули. 
	
	\keywords денатурація ДНК,  модель Поланда-Шераги, полімерні мережі, невпорядковане середовище
\end{abstract}
\label{last@page}
\end{document}